\documentclass[10pt]{article}

\usepackage{epsfig}

\def\lsim{\mathrel{\rlap{\lower4pt\hbox{\hskip1pt$\sim$}}
    \raise1pt\hbox{$<$}}}         
\def\gsim{\mathrel{\rlap{\lower4pt\hbox{\hskip1pt$\sim$}}
    \raise1pt\hbox{$>$}}}         

\def\overleftrightarrow#1{\vbox{\ialign{##\crcr
    $\leftrightarrow$\crcr
    \noalign{\kern 1pt\nointerlineskip}
    $\hfil\displaystyle{#1}\hfil$\crcr}}}

\newcommand{\pp}{\\[0.5cm]}
\newcommand{\be}{\begin{equation}}
\newcommand{\ee}{\end{equation}}
\newcommand{\bea}{\begin{eqnarray}}
\newcommand{\eea}{\end{eqnarray}}

\newcommand{\vk}{k}
\newcommand{\ak}{k}

\begin{document}

\hfill {\bf TUM/T39-02-03} \\
\vspace{0.01in}
\hfill {May 2002} \\

\vspace{0.25in}

\begin{center}
{\bf \Large Debye Screening at Finite \\ Temperature, Revisited\footnote{Work
supported in part by BMBF and GSI.}}
\end{center}
\vspace{0.25 in}
\begin{center}
{R.A. Schneider}\\
{\small \em  Physik-Department, Technische Universit\"{a}t M\"{u}nchen \\D-85747 Garching,
GERMANY\\}
%
\end{center}
\vspace{0.25 in}

\begin{abstract}
We present an alternative way  to calculate the screening of the static potential between two charges
in (non)abelian gauge theories at high temperatures. Instead of a  loop expansion of a gauge boson
self-energy, we evaluate the energy shift of the vacuum to order $e^2$ after applying an external
static magnetic field and extract a temperature- and momentum-dependent dielectric permittivity. The
Hard Thermal Loop (HTL) gluon and photon Debye masses are recovered from the lowest lying Landau levels
of the perturbed vacuum. In QED, the complete calculation exhibits an interesting cancellation of
terms, resulting in a logarithmic running $\alpha(T)$. In QCD, a Landau pole in $\alpha_s$ arises in
the infrared from the sign of the gluon contribution, as in more sophisticated thermal renormalization
group calculations.
\end{abstract}
\vspace{0.25in}
\section{Introduction}
In quantum field theory, fluctuations of the vacuum give rise to the production of pair quanta which tend to screen (or antiscreen) the charge of a heavy test particle. If one perturbatively calculates the non-relativistic potential $V(r)$ between two unlike static charges, say, in QED, the usual Coulomb-like behaviour is modified by the photon self-energy $\Pi(K^2)$ such that
\be
V(r) = \int \frac{d^3 k}{(2\pi)^3} \ e^{i\vec{k} \cdot \vec{r}} \frac{-e^2}{\ak^2 + \Pi(K^2 = -\ak^2)},
\ee
where $k = |\vec{k}|$ and $K = (k^0, \vec{k})$. Inserting the text-book result for $\Pi(K^2)$ and expanding for small distances $\ak^2 \gg m_e^2$, the quantum
fluctuations lead to an effective coupling constant
\be
\alpha_{\rm eff}(\vk) = \frac{\alpha}{\displaystyle 1 - \frac{\alpha}{3\pi} \log\left(\frac{\ak^2}{\Lambda^2} \right)},
\label{alpha_QED}
\ee
where $\Lambda = \exp(5/3) m_e$ is a scale related to the electron mass $m_e$. This is, of course, the
familiar result of the running coupling in QED which is commonly obtained using renormalization group
methods. In refs.\cite{NKN, JLP}, it has been shown that the running of a coupling constant at $T=0$
can be understood in physical terms by the polarizability of the vacuum. The effects of fluctuations
can be incorporated to a certain extent in a scale dependent dielectric permittivity $\epsilon$ that
defines an effective charge $\alpha_{\rm eff} = \alpha/\epsilon$. In vacuum, Lorentz invariance
dictates that
\begin{equation}
\mu \epsilon = 1, \label{epsmu}
\end{equation}
where $\mu$ is the  magnetic permeability. Calculating  $\mu(\vk)$ at the momentum scale $\ak$ and
extracting the leading log contribution, one finally recovers the familiar expressions for the running
couplings in QED and QCD. Then, asymptotic freedom can be interpreted in terms of a paramagnetic ground state.
\pp
In this work, we extend the approach of \cite{NKN, JLP} to finite temperature and calculate an
effective coupling constant $\alpha_{\rm eff}(\vk, T)$.   Instead of a loop expansion, we evaluate the
energy shift of the vacuum to order $e^2$ after applying an external (chromo)magnetic field $H$. The
connection of magnetic permeability and dielectric permittivity at finite temperature is made by
invoking a renormalization group argument. QCD with a magnetic background field at finite temperature
has been studied in a number of works \cite{ANN}. In contrast to previous approaches, we lay out a
non-technical calculation of charge screening without reference to propagators or self-energies,
resorting to entities that have an immediate physical interpretation (energy densities and
susceptibilities). Our work then allows an alternative, though slightly more phenomenological, view on
screening at high temperature.
\pp
A possible dissolution of bound quarkonia states, e.g. $J/\psi$, was proposed long ago as an experimental signature of the quark-gluon plasma in heavy-ion
collisions. The $T$-dependence of the interquark potential in QCD is therefore of
great interest, and simulations of the potential in lattice gauge theory do indeed show a strong screening \cite{LATTICE_POT}. In perturbation theory, the quantity that enters the Fourier transform of the potential
at finite temperature is the static limit of the longitudinal gauge boson self-energy $\Pi_L(k^0,
\vk; T)$ \cite{LB}:
\be
V(r, T) = \int \frac{d^3 k}{(2\pi)^3} \ e^{i\vec{k} \cdot \vec{r}} \frac{-e^2}{\ak^2 + \Pi_L(0, \vk; T)}.
\label{Pot_T}
\ee
Equivalently, one can define a dielectric permittivity by \cite{HAW}
\be
\epsilon(\vk, T) = 1 + \frac{\Pi_L(0, \ak, T)}{\vk^2}. \label{epsilon_def}
\ee
The perturbative one-loop thermal contribution to $\Pi_L$ has been calculated long ago as \cite{ES}:
\be
\Pi_L(0, \vk, T) = \frac{e^2 T^2}{3} \equiv  (m_D^{e})^2 \quad \mbox {for QED,} \label{m_D_QED}
\ee
and
\be
\Pi_L(0, \vk, T) = \left( N_c + \frac{N_f}{2} \right) \frac{g^2 T^2}{3} \equiv (m_D^{c})^2 \quad \mbox {for
QCD,} \label{m_D_QCD}
\ee
which defines screening masses $m_D$. Here, $e$ and $g$ are the electromagnetic and strong coupling,
respectively, $N_c$ is the number of colours and $N_f$ the number of thermally active flavours. Since
the static limits of the self-energies are momentum-independent, the poles of the expression in
(\ref{Pot_T}) are simply the gauge invariant Debye masses $m_D$ defined in eqs.(\ref{m_D_QED}) and
(\ref{m_D_QCD}) and lead to an exponential damping of the potential $V(r) \sim \exp(-m_D r) / r$. In
particular, this form of $\Pi_L$ has the consequence that gluons {\em screen} the strong interaction,
in contrast to the zero temperature case, over long distances. However, the formula for the running QCD
coupling constant commonly used in finite temperature calculations assumes that typical momentum
transfers are of the order of the temperature, hence
\be
\alpha_s(T) \sim \frac{1}{(11 N_c - 2 N_f) \log(T)}.
\ee
In this expression, gluons therefore retain their antiscreening property, reflecting the ultraviolet sector of the theory.
The transition to Debye screening is not obvious. Another troublesome feature of QCD screening at
finite temperature is the behaviour of the Debye mass at next-to-leading order, which reads \cite{AKR}
\be
m_D^{(2)}(T) = m_D^{(1)} + \frac{N_c g^2 T}{4\pi} \log\left(\frac{C}{g} \right), \label{NLO_MD}
\ee
with $m_D^{(1)}$ given by eq.(\ref{m_D_QCD}). Here $C$ is a constant arising from the {\em ad
hoc} removal of infrared singularities involving chromomagnetic static modes. The appearance of the non-perturbative logarithmic term questions the applicability of loop calculations somewhat. Furthermore, whereas in QED the self-energy tensor
$\Pi_{\mu\nu}$ is gauge independent, this is not the case in QCD, which makes the very
definition of a Debye mass conceptually difficult. Finally, due to the nonlinear coupling of the gluons, relation (\ref{epsilon_def}) remains valid only within certain gauges (like temporal axial gauge) \cite{kapusta}. An evaluation of the effective charge and its possible screening not relying on a Feynman graph expansion is therefore desirable.

%
%
\section{The zero temperature case}
%
%
In this section, we define our notation and briefly review the calculation of refs.\cite{NKN, JLP}. To
obtain a scale-dependent $\mu$, let us look at the change in the energy $E$ of the vacuum when an
external magnetic field $H$ is applied:
\be
\Delta E = - \frac{1}{2} \left[ 4\pi \chi(H) \right] VH^2 - E_{\rm vac}, \label{master1}
\ee
where $\mu(H) = 1 + 4\pi \chi(H),$ and $\chi(H)$ is the field-dependent magnetic susceptibility. As soon
as the energies at $H = 0$ and finite $H$ are known to some approximation, a field-, or equivalently,
scale-dependent $\mu(H)$ can be extracted. Later, the external field $2eH$ is identified with the scale
$K^2$ at which the physical process is probed.
\pp
For charged scalar fields, the general expression for the energy spectrum of a single Fourier mode reads
\be
E^\pm_{n, \vk} = \omega_{\vk} \left(n^\pm_{\vk} + \frac{1}{2}\right), \label{E_general}
\ee
distinguishing between particles (+) and antiparticles ($-$). The dispersion relation $\omega_{\vk} =
\ak$ follows from the positive energy solution of the Klein-Gordon equation for massless, non-interacting particles. At
$T=0$, the occupation number $n^\pm_{\vk}$ for the ground state is zero. Summing over particle and antiparticle
states, we recover the familiar divergent zero-point vacuum energy $E_0 = \sum_k \omega_{\vk}$. For
massless spin-$\frac{1}{2}$ fermions, the energy without an external field becomes
\begin{equation}
E_0^f = - 2 \sum_{k} \omega_{\vk}.
\end{equation}
The factor 2 arises from the spin summation, the factor $-1$ stems from the anti-commutation relation
fermionic annihilation and creation operators obey. In the presence of the magnetic field $H$, we
substitute $\partial_\mu \rightarrow D_\mu = \partial_\mu - igq  A_\mu$, where $q$ is the charge of the
(anti)particle in units of the coupling $g$. Choosing the orientation of the $H$-field along the
$z$-axis, we construct a vector potential as $A_\mu = (0,0,x_1 H,0)$. This choice for $A_\mu$ obeys
$\partial_\mu A^\mu = 0$. In the following, we treat QED and QCD in parallel and define $e = qg$.  We have to solve for the energy
spectrum of $i /\!\!\!\!D \psi(x) =0$, which is basically a relativistic version of the Landau theory
for the diamagnetic properties of an electron gas. The solution for the energy of a single Fourier mode becomes
\be
E_{n,k_3,s_3} = \sqrt{k_3^2 + 2eH \left(n + 1/2 + s_3 \right)}. \nonumber
\ee
In addition, the $x_1$ space variable is shifted by $ - k_2/(eH)$. Note that the energy depends only on
two quantum numbers. The third is 'hidden' in the mentioned $x_1$ shift. Here $s_3 = \pm \frac{1}{2}$,
the $z$-component of the spin. The $H s_3$ term clearly shows the coupling of the spin to the external
field, and hence, if the spin of the fermion is anti-parallel to the $H$-field, the energy is lowered. For QCD,
there is also an implicit sum over the colour charges $q$ hidden in $e = gq$. Finally, for a vector gauge boson the $H$-independent energy is the same as for a scalar field, except that
there is an additional factor of 2 counting the transverse spin degrees of freedom:
\begin{equation}
E_0^g = 2 \sum_{k} \omega_{\vk}.
\end{equation}
The sum over colour degrees of freedom yields an additional multiplicative factor of $N_c^2 -1$. In
presence of the magnetic field, we separate the field $A_\mu$ into the classical background
part $A_\mu^{b}$ and the fluctuating quantum part $A_\mu^q$. The equations of motion become  $D_\mu
G^{\mu\nu} = 0$, where $G^{\mu\nu}$ is the gluon field strength tensor. With a suitable choice of background
gauge, the energy for the two physical degrees of freedom of $A_\mu$ can be written
as
\be
E_{n,k_3, s_3} = \sqrt{k_3^2 + 2eH \left( n + 1/2 + s_3 \right)}, \label{E_gluon}
\ee
the same as in the fermionic case, but now with $s_3 = \pm 1$. Again, summation over the colour charges is
implicitly assumed.
\pp
We want to extract the leading log($H$) contribution to the energy shift induced by the external field.
With the total spin $s$ of the particle considered and $i = f,g$:
\begin{eqnarray}
\Delta E^i & = & (-1)^{2s} \left( \sum_{n,k_2,k_3,s_3} E^i_{n,k_3,s_3} - \sum_{k_1,k_2,k_3, s_3}
\omega_{\vk}\right), \quad \mbox{where} \label{e1}
\\ E^i_{n,k_3,s_3} & = & \sqrt{k_3^2 + 2eH \left( n + 1/2 + s_3 \right)}. \label{e2}
\end{eqnarray}
Introducing a quantization volume $V = L^3$, we replace the sum over $k_2$ and $k_3$ by an integral
weighted with the density of states. Taking into account that the $x_1$ variable was shifted, $k_2$ is restricted to $0 \leq k_2 \leq LeH
$. Then,%
\be
\sum_{k_2, k_3} \rightarrow \frac{L}{2\pi} \int dk_3 \ \frac{L}{2\pi} (eH \cdot L) = \frac{V}{4\pi^2} \
(eH) \int dk_3.\nonumber
\ee
To regularize the divergence, we will use a UV cut-off $\Lambda$ such that $0 \leq n \leq
\frac{\Lambda^2}{2eH} = n_\Lambda$ and $k_3^2 \leq \Lambda^2$. The first idea would be to replace the
sum over $n$ by an integral. However, if we perform the shift $n' = 2eH n$, we find that the integral
would be independent of $H$ to leading order. That is, we would have recovered the vacuum result, in
the absence of the field $H$. So what we need is the {\em correction} to the replacement of a sum with
an integral. Such a correction term suitable for our case here is provided by a specific Euler's sum rule
$$
\sum_{n=n_1}^{n_2} f(n+ 1/2) = \int_{n_1}^{n_2} f(x)dx - \frac{1}{24} f'(x) |_{n_1}^{n_2}.
$$
We may now re-define the energy shift as
$$
\Delta E^i = (-1)^{2s} \sum_{s_3} \left\{ \sum_{n=0}^{n_\Lambda} f(n+ 1/2 + s_3) -
\int_0^{n_\Lambda} dn \ f(n+1/2 + s_3) \right\},
$$
where
$$
f(x) = \frac{V}{2\pi^2} (eH) \int_0^\Lambda dk_3 \sqrt{k_3^2 + 2eHx}.
$$
Since we are not interested in the soft modes of the order of $eH$ (the leading-log behaviour is
dominated by the UV behaviour of the theory), we split the sum into two pieces ($N \ll n_\Lambda$)
$$
\sum_{n=0}^{n_\Lambda} = \sum_{n=0}^{N} + \sum_{n=N}^{n_\Lambda}.
$$
Let us treat $s_3$ formally as a continuous variable. Taylor expanding in $s_3$ (since $n \geq N \gg
s_3$), we are left with
\begin{eqnarray}
\Delta E^i & = & (-1)^{2s} \sum_{s_3} \sum_{n=N}^{n_\Lambda} \left( f(n + 1/2) - s_3 f'(n +
1/2) + \frac{s_3^2}{2} f''(n + 1/2) + ... \right)\  + \label{Taylor} \\ & & + \ \Phi(eH,N). \nonumber
\end{eqnarray}
Now $\Phi(eH,N)$, which represents the contributions from soft modes only,  does not depend on $\Lambda$. It is thus proportional to $(eH)^2$ for dimensional reasons, a small non-leading log
contribution, and may be safely neglected. The linear term in $s_3$ vanishes upon summation, and re-substituting $e = gq$, we find
\begin{equation}
\Delta E^i = - \frac{1}{2} V (gH)^2 \left[ \frac{q^2 (-1)^{2s}}{2\pi^2} \sum_{s_3} \left(
\frac{s_3^2}{2} - \frac{1}{24} \right) \log \left(\frac{\Lambda^2}{2eH} \right) \right].\nonumber
\end{equation}
The sum over a $SU(N_c)$ multiplet of the squared charges $q^2$ is $N_f/2$ for the fundamental
representation ($N_f$ quark flavours) and $N_c/2$ for the adjoint representation (the gluons). For QCD,
the susceptibility becomes
$$
4\pi \chi \rightarrow  - g^2 \frac{11N_c - 2N_f}{48\pi^2} \ \log \left(
\frac{2eH}{\Lambda^2} \right),
$$ which reproduces the expression obtained by renormalization group calculations if we identify $2eH =
K^2$. For QED, the sum over the charge(s) is simply 1, so we obtain
$$
4\pi \chi \rightarrow + \ \frac{e^2}{12\pi^2} \log \left( \frac{2eH}{\Lambda^2} \right),
$$
again in accordance with eq.(\ref{alpha_QED}). Having outlined the calculation of \cite{NKN, JLP}, we
now proceed to the main part of the paper and switch on temperature.
%
%
\section{The temperature-dependent part}
%
%
At finite temperature $T$, the occupation number $n_\vk^\pm$ appearing in eq.(\ref{E_general}) does not
vanish anymore for the thermal ground state, instead $n_{\vk} = (\exp(\beta \omega_{\vk}) -1 )^{-1} = n_{BE}$, the usual
Bose-Einstein distribution function ($\beta = 1/T$). For fermions, $n_\vk = (\exp(\beta \omega_{\vk}) + 1 )^{-1} = n_{FD}$, the Fermi-Dirac distribution
function. Thus, when summing over the infinitely many
degrees of freedom, we find for the total vacuum energy of a charged scalar field
\begin{equation}
E_0^s = \sum_{k} \omega_{\vk} \left( 1 + 2 n_{BE}(\omega_{\vk}) \right).\label{scalarH0}
\end{equation}
The result clearly separates into the divergent vacuum part already treated and a finite, $T$-dependent part.
 In the case of a finite magnetic field $H$, the higher energy modes (\ref{e2}) are occupied with their respective thermal probabilities, and we can write ($i = f,g$):
\begin{eqnarray}
\Delta E^i & = & E^i - E^i_0, \label{dE}\\
E^i & = & \sum_{n,k_2,k_3,s_3} E^i_{n,k_3,s_3} \ \left[ (-1)^{2s} + \frac{2}{\exp(\beta E^i_{n,k_2,s_3}) -
(-1)^{2s}} \right]  \quad \mbox{and} \label{e3} \\
E_0^i & = & \sum_{k_1,k_2,k_3,s_3} \omega_\vk \ \left[ (-1)^{2s} + \frac{2}{\exp(\beta \omega_\vk) -
(-1)^{2s}} \right].
\end{eqnarray}
Again, we need to extract the leading thermal contribution to $\mu(\vk)$. However, at finite
temperature, relation (\ref{epsmu}) does not hold any more. One could imagine to start with
\be
\mu(\vk) \epsilon(\vk) = n(\vk)^2,\label{epsmuT}
\ee
where $n$ is the index of refraction. This quantity is related to the photon or gluon phase velocity by
$v_p = 1/n$, and $v_p$ could be extracted from the (full) dispersion relation of the corresponding
gauge boson since $v_p = \omega_\vk/\ak$. However, eq.(\ref{epsmuT}) holds only for "on-shell"
propagating gauge bosons. Since Lorentz invariance is formally broken by the presence of the heat bath,
$\mu$ and $\epsilon$ become functions of $k^0$ and $k$, and eq.(\ref{epsmuT}) reads, more explicitly,
$\mu(\omega_k, k) \epsilon(\omega_k, k) = n(k)^2$. Here, we consider an off-shell external field, so a
relation between $\mu(0, k)$ and $\epsilon(0, k)$ is required.
\pp
The total energy density of the system can be written as the sum of the field
 and the induced medium energy density:
\be
\mathcal{E}_{\rm tot} = \frac{1}{2} H^2 + \frac{1}{V} \sum \limits_{i} \Delta E^i. \label{E_tot}
\ee
Introducing an effective field $H_{\rm eff}$, we rewrite $\mathcal{E}$ as
$$
\mathcal{E}_{\rm tot} = \frac{1}{2} H_{\rm eff}^2 = \frac{1}{2} \frac{(eH)^2}{e_{\rm eff}^2}.
$$
In the last step we made use of the fact that $eH$ has to be a renormalization group invariant, so $eH
= e_{\rm eff} H_{\rm eff}$. The effective coupling constant is now defined by \cite{JLP, ANN}
$$
\frac{1}{e_{\rm eff}^2} \equiv 2 \frac{\partial \mathcal{E}_{\rm tot}}{\partial (eH)^2} = \frac{1}{e^2}
[1 - 4\pi \chi(2eH, T) ],
$$
using (\ref{master1}) and (\ref{E_tot}). Replacing $2eH$ by $k^2$, as at $T=0$, our master formula
hence reads
\be
\alpha_{\rm eff}(\vk, T) = \frac{\alpha}{\epsilon(\vk, T)} = \frac{\alpha}{1 - 4\pi \chi(\vk,
T)}.\label{eps_mu_T}
\ee
The thermal piece of eq.(\ref{e3}) can be compactly re-written as
\begin{equation}
E^{th}(T, b, s, s_3) = V T^4 \left(\frac{b}{2\pi^2} \sum_{s_3} \sum_{n=0}^{\infty} \int \limits_0^\infty dx \
\frac{\sqrt{x^2 + b(n + 1/2 + s_3)}}{\exp\left(\sqrt{x^2 + b(n+1/2 + s_3)}\right)  - (-1)^{2s} }
\right), \label{e1th}
\end{equation}
where $x$ is dimensionless and $b = 2eH/T^2$ is a measure for the ratio of quantum and thermal effects. We consider the high-temperature limit $b \ll 1$ in the rest of the paper.
%
%
\subsection{A first (incomplete) approximation}
%
%
The sum appearing in expression (\ref{e1th}) obviously cannot be evaluated exactly. It is instructive to work out the first intuitive approximation to the sum although we will show in the next section that it is too crude.
\pp
Consider the fermionic part.  Note that the factor $b(n + 1/2 + s_3)$ plays the role of a mass term in
the integral in eq.(\ref{e1th}), so the contribution of the terms in the sum becomes exponentially
suppressed as $n$ increases. In contrast to the $T=0$ case we are therefore interested in the behaviour
of the sum at {\em small} $n$ where the $s_3$ spin component is not negligible. Thus we cannot apply a
Taylor expansion in $s_3$, as done in (\ref{Taylor}), and need an exact summation over $s_3$. Isolating
then the lowest lying Landau mode ($n=0, s_3 = - 1/2$) and combining the remaining expressions into a
single sum, we find
\be
\frac{E^f}{VT^4} \equiv \tilde{E}^f = \frac{b}{24} + \frac{b}{\pi^2} \sum \limits_{n=0}^\infty \int \limits_0^\infty dx \frac{\sqrt{x^2 + b(n+1)}}{\exp\left( \sqrt{x^2 + b(n+1)} \right) + 1}.\label{E_f}
\ee
Since $b \ll 1$, the terms in the sum vary slowly with $n$, so we can again try to trade the sum for
an integral over $n$:
\be
\tilde{E}^f =  \frac{b}{24} +  \frac{2}{\pi^2} \int \limits_0^\infty dr \frac{r^2 \sqrt{r^2 + b}
}{\exp\left( \sqrt{r^2 + b} \right) + 1}. \label{E_f_1}
\ee
Neglecting terms of order $b$ in the integral, we obtain as a first approximation
\be
E_{(0)}^{f}  =  - \frac{1}{2} V H^2 \left[ - \frac{(m_D^e)^2}{2eH} \right] + \frac{7}{4} \ \frac{\pi^2}{15}  VT^4 \label{E_f_simple_QED}
\ee
for QED with the HTL Debye mass defined in eq.(\ref{m_D_QED}). The second term is simply the energy
$E_0^f(T)$ of a thermally excited, non-interacting massless fermion-antifermion pair, i.e. the thermal
energy of the {\em unperturbed} vacuum that has to be subtracted, cf. eq.(\ref{dE}). This means that we
have recovered within our simple framework the perturbative one-loop HTL result from the lowest Landau
level contribution to the energy of the magnetically perturbed thermal vacuum. The energy difference
that enters in (\ref{master1}) already yields $4\pi\chi(H, T)$ as the expression in square brackets,
and the effective coupling constant reads, following eq.(\ref{eps_mu_T}),
$$
\alpha_{\rm eff}(\vk, T) = \frac{\alpha}{\epsilon(\vk, T)} = \frac{\alpha}{\displaystyle 1 +
 \frac{(m_D^e)^2}{\ak^2}},
$$
as within HTL perturbation theory, eq.(\ref{epsilon_def}).
\pp
For QCD with $N_f$ flavours, we obtain
\be
E_{(0)}^{f}  =  - \frac{1}{2} V H^2 \left[ - \frac{m_{D,f}^2}{2eH} \right] + N_f \frac{7 \pi^2}{60} \ VT^4 \label{E_f_simple}
\ee
with the fermionic part of the squared QCD Debye mass (\ref{m_D_QCD}), $m_{D,f}^2 = N_f/6 \ g^2 T^2$.
For the total evaluation of the QCD susceptibility, we need to add the contribution from the gauge
bosons. At zero temperature, contributions from ``unphysical'' gluon states in the calculation of the
energy spectrum, eq.(\ref{E_gluon}), are exactly cancelled by Fadeev-Popov ghost contributions within
the background gauge condition used here. Since we only consider excitations of energy levels that were
evaluated at $T=0$, no ambiguity arises and we still work only with physical gluon degrees of freedom
with two polarization states. We proceed in close analogy to the fermionic case: first, we sum over
$s_3 = \pm 1$. A subtlety arises since the combination $n=0$ and $s_3 = -1$ in eq.(\ref{e1th}) issues a
negative value under the square root for small $x$. This 'tachyonic'  mode is related to a possible
instability of the vacuum \cite{TACH}. Its effect on the magnetic field over large distances, however,
is negligible, the sum over $n$ for $s_3 = -1$ therefore starts only at $n=1$. Isolating again the
lowest lying physical Landau level ($n=1, s_3 = -1$) contribution to the sum, we are left with
\be
\frac{E^g}{VT^4} \equiv \tilde{E}^g = \frac{b}{2 \pi^2} \int \limits_0^\infty dx \frac{\sqrt{x^2 +
b/2)}}{\exp\left( \sqrt{x^2 + b/2} \right) -1} + \frac{b}{\pi^2} \sum \limits_{n=0}^\infty \int
\limits_0^\infty dx \frac{\sqrt{x^2 + b(n+3/2)}}{\exp\left( \sqrt{x^2 + b(n+3/2)} \right) -
1}.\label{E_g}
\ee
Replacing the sum by an integration, setting $b = 0$ in the integrals and summing over colour, the
result becomes
\be
E^g_{(0)} = - \frac{1}{2} V H^2 \left[- \frac{m_{D,g}^2}{2eH}  \right] +  2 (N_c^2 - 1) \frac{\pi^2}{15} VT^4.\label{E_g_simple}
\ee
Again, the last term is the thermal energy $E_0^g(T)$ of the  unperturbed SU(N$_c$) gluon vacuum. The
expression in square brackets exactly corresponds to the gluonic part of the squared QCD Debye mass,
$m_{D, g}^2 = N_c/3 \ g^2 T^2$. Putting all pieces together, the effective coupling becomes
$$
\alpha_{s, \rm eff}(\vk, T) = \frac{\alpha_s}{\displaystyle 1 +  \frac{(m_D^c)^2}{\ak^2}},
$$
very similar to the QED case. In our model, the Hard Thermal Loop (chromo)electric Debye masses
therefore appear as the lowest Landau level contribution to the energy difference that arises when one
probes the thermal vacuum by a (chromo)magnetic field. It is worth noting that, in this approximation,
the alignment of an external field always {\em increases} the thermal energy of the vacuum, regardless
of the non-abelian structure of the theory. Therefore $\chi(\vk, T)$ is always negative and we
conclude, using eq.(\ref{eps_mu_T}), that the static potential would become screened by both fermions
and by gauge bosons.
%
%
\subsection{A better approximation}
%
%
However, additional contributions to eqs.(\ref{E_f_simple_QED}),
(\ref{E_f_simple}) and (\ref{E_g_simple}) of the {\em same order in $e^2$} arise from two sources.
First, the expansion of the integrals (\ref{E_f_1}) and (\ref{E_g}) in $b$ is similar to the high temperature expansion of loop integrals with massive particles in the small quantity $m_0/T$. The appendix contains the relevant formulae. Second, the correction to the replacement of the sum by an integral yields terms to order $b$ and $b^2$ that are provided by the Euler-MacLaurin formula
\be
\sum \limits_{n=0}^N f(n) = \int \limits_0^N f(x) dx + \frac{1}{2} \left[ f(N) + f(0) \right] + \frac{1}{12} \left[ f'(N) - f'(0) \right] + \dots , \label{Euler}
\ee
where the dots denote terms with higher derivatives in $f(n)$. For our purposes, eq.(\ref{Euler}), taking $N \rightarrow \infty$, is sufficient, as long as $f(x) \in  \mathcal{C}^2$ for $x \in [0,N]$. When calculating the thermal contribution to the vacuum energy, we include these correction terms to the integral in the following and expand all integrals in the small parameter $b$, using the relations presented in the appendix. The summation of all terms to order $e^2$  then alters the results in (\ref{E_f_simple}) and (\ref{E_g_simple}) {\em qualitatively}.
%
\subsection{Results for QED}
%
For fermions, we start with eq.(\ref{E_f}). Defining $\delta^2 = b$, we obtain to order $\delta^4$
\bea
\tilde{E}_{(1)}^{f} & = & \frac{\delta^2}{24} +  \frac{2}{\pi^2} \int \limits_0^\infty dr \frac{r^2
\sqrt{r^2 + \delta^2}}{\exp(\sqrt{r^2 + \delta^2}) + 1} + \frac{1}{2} \left[
\frac{\delta^2}{\pi^2} \int \limits_0^\infty dr \frac{\sqrt{r^2 + \delta^2}}{\exp(\sqrt{r^2 +
\delta^2}) + 1} \right] + \nonumber \\ & & + \frac{1}{12} \left[ - \frac{\delta^4}{2\pi^2} \int \limits_0^\infty dx \left\{ \frac{1}{\sqrt{x^2 + \delta^2}} - \frac{1}{1 + \exp(-\sqrt{x^2 + \delta^2})} \right\} \frac{1}{\exp(\sqrt{x^2 + \delta^2}) + 1} \right]
%
\eea
Using the functions $f_i(y)$ and $g^+(y)$ defined in the appendix, we re-write
\be
\tilde{E}_{(1)}^{f}  =  \frac{\delta^2}{24} +  \frac{2}{\pi^2} f_5(\delta) + \frac{5\delta^2}{2\pi^2} f_3(\delta) + \frac{\delta^4}{2\pi^2} f_1(\delta) - \frac{\delta^4}{24\pi^2} g^+(\delta).\label{f_func}
\ee
Expanding in $\delta$ and keeping all terms up to $\mathcal{O}(\delta^4)$, surprisingly all terms
of order $\delta^2$ {\em cancel}, and we are left with
\be
\tilde{E}_{(1)}^{f} = \frac{7\pi^2}{60} + \frac{\delta^4}{96 \pi^2} \log\left(\mathcal{A}_f \delta^2 \right)
\ee
with  $\gamma = 0.5772...$ and the constant  $\mathcal{A}_f = \exp(2\gamma -1)/\pi^2 \simeq 0.12$. The
first term is the well-known thermal part of the vacuum energy in the absence of the field $H$. Since
$\delta^2 \ll 1$, the alignment of a magnetic field hence {\em decreases} the energy of the vacuum at
finite temperature, in contrast to the result of the previous section. The susceptibility in QED
therefore becomes
\be
4 \pi \chi(H, T) = - \frac{\alpha}{3\pi} \log \left(\mathcal{A}_f \frac{2eH}{T^2}\right). \label{chi_QED}
\ee
Note that the pre-factor of the logarithm is the same  as in the zero-temperature case! Including this
pure quantum correction for the QED running coupling, eq.(\ref{alpha_QED}), all momentum dependence
drops out, and we finally obtain with (\ref{eps_mu_T})
\be
\alpha_{\rm eff}(\vk, T) = \frac{\alpha}{\displaystyle 1 - \frac{\alpha}{3\pi} \log \left(
\frac{T^2}{\Lambda_T^2} \right)} \quad \mbox{with } \Lambda_T = m_e \frac{e^{\gamma + 7/6}}{\pi} \simeq
\frac{\Lambda}{3},
\ee
which is valid to order $e^2$ and for momenta
\be
m_e \ll \ak \ll T. \label{kinematic}
\ee
Using the zero-temperature coupling $\alpha_{\rm eff}(\vk)$ from eq.(\ref{alpha_QED}), the effective
coupling can, to this order, be rewritten as
\be
\alpha_{\rm eff}(\vk, T) = \alpha_{\rm eff}(T) = \alpha_{\rm eff}(\langle \vk \rangle \simeq 3 \ T).
\ee
So the common practice used in perturbation theory to simply take the running of the coupling at zero
temperature and set as the scale the thermally averaged momentum scale $\langle k \rangle \simeq 3 \ T$
does indeed find support from our calculation for QED.
%
\subsection{Results for QCD}
%
For QCD, the fermionic contribution takes a form similar to the QED result,
\be
4 \pi \chi(H, T) = - \frac{g^2 N_f}{24\pi^2} \log \left(\mathcal{A}_f \frac{2eH}{T^2}\right).\label{chi_QCD}
\ee
Note that the pre-factor of the logarithm is again the same as at zero temperature. The calculation of the gluonic
part of $\chi$ runs along the same lines. Starting with eq.(\ref{E_g}) and setting $\delta^2 = b/2$, we
obtain by use of the functions $h_i(y)$ and $g^-(y)$
\be
\tilde{E}^g = \frac{2}{\pi^2} h_5(\sqrt{3} \delta) + \frac{\delta^2}{\pi^2} \left[ 7 h_3(\sqrt{3} \delta) + h_3(\delta) \right] + \frac{\delta^4}{\pi^2} \left[h_1(\delta) + 3 h_1(\sqrt{3} \delta) - \frac{1}{6} g^-(\sqrt{3} \delta)  \right]. \label{h_func}
\ee
Expanding, we find that all terms of order $\delta^3$ cancel and the result becomes
\be
\tilde{E}^g = \frac{2 \pi^2}{15} - \frac{b}{12} - \frac{5 b^2 }{128\pi^2} \log \left(\mathcal{A}_g b \right),\label{E_g_4}
\ee
with the constant  $\mathcal{A}_g = \exp(2\gamma - 13/10 + 11/5 \ \log 3)/(32\pi^2) \simeq 0.03$.
Similar to the fermionic part, the alignment of a chromomagnetic field hence always {\em lowers} the
energy of the gluonic vacuum, however, this time the difference goes linearly in $b$ and not only
logarithmically. Finally summing over colour, the gluonic susceptibility reads
\be
4\pi \chi(H, T) = \frac{N_c}{3} \frac{g^2 T^2}{2eH} + 5 N_c \frac{g^2}{48\pi^2} \log \left(\mathcal{A}_g \frac{2eH}{T^2} \right).
\ee
In this expression the sign is reversed as compared to eq.(\ref{chi_QCD}), and $4\pi \chi$ is always
positive. The total thermal result for QCD, excluding the $T=0$ contribution,  becomes
\be
\alpha_{s, \rm eff}(\vk, T) = \frac{\alpha_s}{\displaystyle 1 - \alpha_s \frac{4 \pi N_c}{3}
\frac{T^2}{\ak^2} - \frac{\alpha_s}{12 \pi} \left[5 N_c \log\left(\frac{\mathcal{A}_g \ak^2}{T^2}
\right) - 2N_f \log\left(\frac{\mathcal{A}_f \ak^2}{T^2} \right)  \right] }. \label{QCD_running}
\ee
Since the gluon contribution dominates by far over the logarithmic fermionic, there is {\em
antiscreening} at high temperature and long distances. This result is in contrast to expectation and
lattice results on the interquark potential \cite{LATTICE_POT}. Extrapolating eq.(\ref{QCD_running})
beyond the kinematical region $g \ll \ak/T \ll 1$ where our approximations are valid, a Landau pole
appears in the infrared region $\ak/T \simeq g$. Figure 1 shows the ratio $\alpha_{s, \rm eff}/
\alpha_s$ as a function of $\ak/T$ for a weak coupling $g = 0.1$, compared to the usual HTL result. A
similar behaviour is also found in more sophisticated renormalization group analyses of the running
coupling at finite temperature (see, e.g., \cite{ANN, RG1, RG2, RG3}).  We note that our results
compare quite well with the numerical solutions obtained in ref.\cite{RG2}. Since
eq.(\ref{QCD_running}) depends only on the dimensionless quantity $k/T$, taking the limit $T
\rightarrow \infty$ at large $k$ is in a sense equivalent to probing the infrared region $k \rightarrow
0$ at smaller $T$, indicating that non-perturbative physics plays an important role even at high $T$.
The necessity for a soft magnetic mass $\sim g^2 T$ as an infrared regulator in loop calculations (see
eq.(\ref{NLO_MD})) supports this line of reasoning.
\begin{figure}
\begin{center}
\epsfig{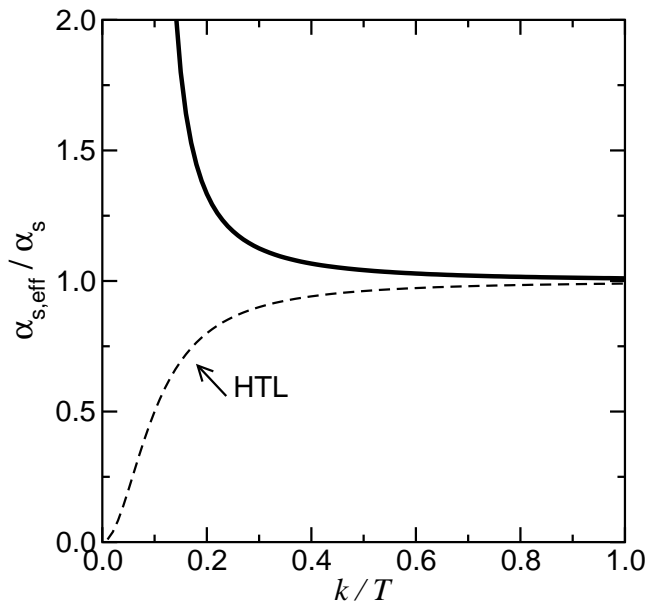}
\caption{The ratio $\alpha_{s, \rm eff}/ \alpha_s$ of eq.(\ref{QCD_running}) as a function of $\ak/T$ for a coupling $g = 0.1$ and $N_f = 0$ (solid line). For comparison, the corresponding ratio in the HTL calculation, using the Debye mass of eq.(\ref{m_D_QCD}), is also plotted (dashed line).}
\end{center}
\end{figure}
\section{Conclusions}
We have presented an alternative way to calculate the screening of the static potential between two
charges in (non)abelian gauge theories at finite temperature by looking at the magnetic properties of
the vacuum. Instead of a loop expansion, we have calculated the energy shift of the vacuum at finite
temperature to order $e^2$ after applying an external (chromo)magnetic field $H$ as a probe. Magnetic
permeability and dielectric permittivity have been connected by a renormalization group argument.

Using a high-temperature expansion $H/T^2 \ll 1$, the gluon and photon Debye masses appearing in the
HTL calculation have been recovered in a first, though incomplete approximation, originating from the
lowest lying Landau level contribution to the thermal energy. Taking into account all contributions to
order $e^2$ in QED, the final expression in the kinematic region (\ref{kinematic}) shows a logarithmic,
momentum-independent running of $\alpha$ with temperature, as expected from simply inserting the
average thermal momentum $\langle k \rangle \simeq 3 \ T$ in the zero-temperature running coupling.

In QCD, we have found indications for a Landau pole at small $\ak/T$ that arises, as in more
sophisticated thermal renormalization group calculations, from the sign of the gluon contribution,
despite well controlled approximations and a completely different approach as compared to conventional
perturbation theory. Our calculation may serve as yet another indication that an expansion in a
presumably small coupling $g$ at high temperatures ceases to yield sensible results for some
quantities, and that this failure is not specific to a Feynman graph expansion. Truly non-perturbative
input that is probably linked to the understanding of confinement is then called for.
\section{Appendix}
Here we present the formulas used to evaluate the $\delta$-dependent integrals appearing in eqs.(\ref{h_func}) and (\ref{f_func}). Fermionic integrals of
the form
$$
f_n(y) = \int \limits_0^\infty dx \ \frac{x^{n-1}}{\sqrt{x^2 + y^2}} \ \frac{1}{\exp(\sqrt{x^2 + y^2}) + 1}
$$
can be expanded for small $y$ (note our slightly different convention compared to \cite{kapusta}).
Especially,
\bea
f_1(y) & = & - \frac{1}{2} \left[ \log\left(\frac{y}{\pi} \right) + \gamma \right] + \dots, \\
f_3(y) & = & \frac{\pi^2}{12} + \frac{y^2}{4} \left[ \log\left(\frac{y}{\pi} \right) + \gamma -
\frac{1}{2} \right] + \dots \quad \mbox{and} \\
f_5(y) & = & \frac{7\pi^4}{120} -  \frac{\pi^2}{8} y^2 - \frac{3}{16} y^4 \left[
\log\left(\frac{y}{\pi} \right) + \gamma - \frac{3}{4} \right] + \dots
\eea
where $\gamma = 0.5772...$ is the Euler-Mascheroni constant. For bosons,
$$
h_n(y) = \int \limits_0^\infty dx \ \frac{x^{n-1}}{\sqrt{x^2 + y^2}} \ \frac{1}{\exp(\sqrt{x^2 + y^2}) - 1}.
$$
The corresponding expansions read
\bea
h_1(y) & = & \frac{\pi}{2y} + \frac{1}{2} \left[ \log\left(\frac{y}{4\pi} \right) + \gamma \right] +
\dots,
\\
h_3(y) & = & \frac{\pi^2}{6} - \frac{\pi}{2} y - \frac{y^2}{4} \left[ \log\left(\frac{y}{4\pi} \right)
+ \gamma - \frac{1}{2} \right] + \dots \quad \mbox{and} \\
h_5(y) & = & \frac{\pi^4}{15} -  \frac{\pi^2}{4} y^2 + \frac{\pi}{2} y^3 + \frac{3}{16} y^4 \left[
\log\left(\frac{y}{4\pi} \right) + \gamma - \frac{3}{4} \right] + \dots
\eea
For the evaluation of derivative terms, we need the leading log($y$) behaviour of integrals such as
\be
g^\pm(y) = g_1^\pm(y) + g_2^\pm(y) = \int \limits_0^\infty dx \left( \frac{1}{\sqrt{x^2 + y^2}} - \frac{1}{1 \pm \exp( - \sqrt{x^2 + y^2})} \right) \frac{1}{\exp(\sqrt{x^2 + y^2}) \pm 1}. \label{formula_deriv}
\ee
The expansion of the first term in brackets, $g_1^\pm$, is known since $g_1^+ = f_1$ and $g_1^- = h_1$. For the evaluation of the second a trick is convenient. Introduce a parameter $\alpha$ to write
\be
g_2^\pm(y; \alpha) = - \int \limits_0^\infty dx \frac{\exp(\alpha \sqrt{x^2 + y^2})}{(\exp(\alpha \sqrt{x^2 + y^2}) \pm 1)^2}. \label{formula1}
\ee
Obviously, $g_2^\pm(y; 1)$ is the sought quantity. Now $g_2^\pm(y; \alpha)$ can also be written as
\be
\frac{\partial}{\partial \alpha} \left[ \int \limits_0^\infty dx \ \frac{1}{\sqrt{x^2 + y^2}} \ \frac{1}{\exp(\alpha \sqrt{x^2 + y^2}) \pm 1} \right] = \frac{d}{d \alpha} g_1^\pm(\alpha y).
\ee
Expanding eq.(\ref{formula1}) for small $y$ hence yields
\bea
g_2^-(y; \alpha) & = & - \frac{1}{\alpha^2} \frac{\pi}{2y} + \frac{1}{2\alpha} + \dots \quad \mbox{for bosons and} \\
g_2^+(y; \alpha) & = & - \frac{1}{2\alpha} + \dots \quad \mbox{for fermions.}
\eea
Setting $\alpha = 1$ and putting the pieces together, the leading-log behaviour of eq.(\ref{formula_deriv}) is
\bea
g^+(y) & = & -  \frac{1}{2} \left[\log\left(\frac{y}{\pi}\right) + \gamma + 1 \right] + \dots \quad \mbox{for fermions and} \\
g^-(y) & =  & +  \frac{1}{2} \left[\log\left(\frac{y}{4\pi}\right) + \gamma + 1 \right] + \dots \quad \mbox{for bosons.}
\eea

\end{document}